\documentclass[twocolumn, tighten]{aastex6}

\usepackage{amsmath}
\usepackage{amstext}
\usepackage{natbib}
\usepackage{graphicx}
\usepackage{acronym}
\usepackage{url}
\usepackage{hyperref}

\usepackage{amssymb}
\usepackage{epsfig}
\usepackage{amsmath}
\usepackage{color}
\usepackage{yfonts}
\usepackage{epsfig}  \usepackage{graphicx}   \usepackage{rotating}

\newcommand{\lsim}{\lower0.6ex\vbox{\hbox{$ \buildrel{\textstyle <}\over{\sim}\ $}}}
\newcommand{\gsim}{\lower0.6ex\vbox{\hbox{$ \buildrel{\textstyle >}\over{\sim}\ $}}}
\newcommand{\beq}{\begin{equation}}
\newcommand{\eeq}{\end{equation}}

\newcommand{\ben}{\begin{enumerate}}
\newcommand{\een}{\end{enumerate}}
\newcommand{\bit}{\begin{itemize}}
\newcommand{\eit}{\end{itemize}}

\newcommand{\rproj}{r_{\rm p}}

\newcommand{\mockobs}{{\tt mock\_observables }}
\newcommand{\emodels}{{\tt empirical\_models }}
\newcommand{\sims}{{\tt sim\_manager }}

\shorttitle{{\tt Halotools} v0.2}
\shortauthors{Hearin, Campbell, Tollerud, et al.}

\begin{document}

\title{Forward Modeling of Large-Scale Structure: \\An open-source approach with Halotools}

\author{Andrew~P.~Hearin\altaffilmark{1,2}, Duncan~Campbell\altaffilmark{3}, Erik~Tollerud\altaffilmark{3,4}\\and\\
Peter~Behroozi\altaffilmark{5,6}, Benedikt~Diemer\altaffilmark{7},
Nathan~J.~Goldbaum\altaffilmark{8}, Elise~Jennings\altaffilmark{9,10}, Alexie~Leauthaud\altaffilmark{11}, Yao-Yuan~Mao\altaffilmark{12,13},
Surhud~More\altaffilmark{11}, John~Parejko\altaffilmark{14}, Manodeep~Sinha\altaffilmark{15,16}, Brigitta~Sip\H{o}cz\altaffilmark{17,18},
Andrew~Zentner\altaffilmark{13}
}
\altaffiltext{1}{Yale Center for Astronomy \& Astrophysics, Yale University, New Haven, CT}
\altaffiltext{2}{Argonne National Laboratory, Argonne, IL, USA 60439, USA}
\altaffiltext{3}{Department of Astronomy, Yale University, P.O. Box 208101, New Haven, CT}
\altaffiltext{4}{Space Telescope Science Institute, Baltimore, MD 21218, USA}
\altaffiltext{5}{Department of Astronomy, University of California, Berkeley, Berkeley CA 94720, USA}
\altaffiltext{6}{Hubble Fellow}
\altaffiltext{7}{Institute for Theory and Computation, Harvard-Smithsonian Center for Astrophysics, 60 Garden St., Cambridge, MA 02138, USA}
\altaffiltext{8}{National Center for Supercomputing Applications, University of Illinois at Urbana-Champaign, 1205 W. Clark St., Urbana, IL 61801, USA}
\altaffiltext{9}{Center for Particle Astrophysics, Fermi National Accelerator Laboratory MS209, P.O. Box 500, Kirk Rd. \& Pine St., Batavia, IL 60510-0500}
\altaffiltext{10}{Kavli Institute for Cosmological Physics, Enrico Fermi Institute, University of Chicago, Chicago, IL 60637}
\altaffiltext{11}{Kavli IPMU (WPI), UTIAS, The University of Tokyo, Kashiwa, Chiba 277-8583, Japan}
\altaffiltext{12}{Kavli Institute for Particle Astrophysics and Cosmology and Department of Physics, Stanford University, Stanford, CA 94305, USA}
\altaffiltext{13}{Department of Physics and Astronomy and Pittsburgh Particle Physics, Astrophysics, and Cosmology Center (PITT PACC),\\{}University of Pittsburgh, Pittsburgh, PA 15260, USA}
\altaffiltext{14}{University of Washington, Dept. of Astronomy, Box 351580, Seattle, WA 98195}
\altaffiltext{15}{Department of Physics \& Astronomy, 6301 Stevenson Center, Vanderbilt University, Nashville, TN 37235, USA}
\altaffiltext{16}{Centre for Astrophysics \& Supercomputing, Swinburne University of Technology, 1 Alfred St, Hawthorn, VIC 3122, Australia}
\altaffiltext{17}{Centre for Astrophysics Research, University of Hertfordshire,College Lane, Hatfield, AL10 9AB, UK}
\altaffiltext{18}{Institute of Astronomy, University of Cambridge, Madingley Road, Cambridge, CB3 0HA, UK}


\begin{abstract}
We  present the first stable release of {\tt Halotools} (v0.2), a community-driven Python package designed to build and test models of the galaxy--halo connection. {\tt Halotools} provides a modular platform for creating mock universes of galaxies starting from a catalog of dark matter halos obtained from a cosmological simulation. The package supports many of the common forms used to describe galaxy--halo models: the halo occupation distribution (HOD), the conditional luminosity function (CLF), abundance matching, and alternatives to these models that include effects such as environmental quenching or variable galaxy assembly bias. Satellite galaxies can be modeled to live in subhalos, or to follow custom number density profiles within their halos, including spatial and/or velocity bias with respect to the dark matter profile. The package has an optimized toolkit to make mock observations on a synthetic galaxy population, including galaxy clustering, galaxy--galaxy lensing, galaxy group identification, RSD multipoles, void statistics, pairwise velocities and others, allowing direct comparison to observations. {\tt Halotools} is object-oriented, enabling complex models to be built from a set of simple, interchangeable components, including those of your own creation. {\tt Halotools} has an automated testing suite and is exhaustively documented on \url{http://halotools.readthedocs.io}, which includes quickstart guides, source code notes and a large collection of tutorials. The documentation is effectively an online textbook on how to build and study empirical models of galaxy formation with Python.
\end{abstract}
\maketitle

\section{Introduction}
\label{section:introduction}

Empirical modeling of the galaxy--halo connection has become a mature subfield of galaxy formation studies. The foundation of this subfield is the halo model \citep{mafry00,seljak00}, which gives an approximate description of the large-scale density field by supposing that all matter is bound within collapsed regions called {\em dark matter halos} \citep[see][for reviews]{cooray02,mo_vdb_white10}. The abundance, internal structure and spatial distribution of dark matter halos can be predicted via simulations with high precision \citep[e.g.,][]{tinker08a,tinker10,bhattacharya_etal11,bhattacharya_etal13,diemer_kravtsov15,heitmann_etal16}. Since the centers of dark matter halos are the natural sites of galaxy formation \citep{whiterees78}, then quantitative predictions for the statistical distribution of galaxies are enabled by knowledge of how galaxy properties are connected to the properties of their underlying halos. 

The modern methods of galaxy--halo modeling has been in place for over ten years, including formulations such as the Halo Occupation Distribution \citep[HOD, ][]{berlind02}, the Conditional Luminosity Function \citep[CLF,][]{yang03} and abundance matching \citep{kravtsov04a,vale_ostriker04,conroy06,vale_ostriker06}. Because of the computational efficiency and the transparency of the assumptions underlying empirical models, these are the most widely used approaches to incorporating galaxy formation physics into contemporary cosmological likelihood analyses \citep[e.g.,][]{cacciato_etal13, reid_etal14}. Galaxy--halo models also play a central role in uncovering observational trends that are now considered fundamental to extragalactic astronomy, such as the stellar mass-to-halo mass relation \citep{tinker05, vdBosch07, behroozi10, moster10, leauthaud_etal12} and the demographics of galaxy quenching \citep{vdBosch03a, collister05, behroozi13b, tinker_etal13}.

Building upon now-standard formulations of the galaxy--halo connection, recent advances in this field have seen an increase in the complexity of empirical models. For example, in the abundance matching formulation constrained in \citet{lehmann_etal15}, stellar mass is influenced by both present-day halo mass as well as halo concentration. Satellite galaxies in the ``delayed-then-rapid" model introduced in \citet{wetzel_etal12b} have an explicit dependence on the time the satellites first passed within the virial radius of their host halo. In the age matching model \citep{HW13a}, the star-formation history of both centrals and satellites is tightly coupled to halo mass assembly history across cosmic time. These advances reflect a growing trend in galaxy--halo modeling that we only expect to continue: galaxies {\em co-evolve} together with their parent halos. While this improves the sophistication of the predictions that can be extracted from empirical models, it also creates additional technical challenges and hurdles for their implementation, especially for beginners.

The standardization of traditional models such as the HOD, CLF and abundance matching provides a natural motivation for a correspondingly standard code base with which the observable predictions of these models can be generated. The trend of galaxy--halo models towards increasing complexity highlights the need for such a code base, particularly one that is open-source and transparent to modify and extend.

Alongside these scientific developments, over the past several years the Python programming language\footnote{\url{http://www.python.org}} has become the most popular language in the field of astronomy \citep{momcheva_tollerud15}. Many aspects of Python make this language well-suited for a fast-moving scientific field: dynamic typing and automated memory management facilitate rapid development; stable, high-performance libraries such as {\tt NumPy}\footnote{\url{http://www.numpy.org}} \citep{numpy_array} and {\tt SciPy}\footnote{\url{https://www.scipy.org}} \citep{scipy} relieve scientists from the need to implement common numerical routines in scientific computing; the Python community has developed a broad toolkit for transparently documenting and testing code in an automated fashion; Python users find its uncluttered syntax to be highly readable and expressive. A significant additional factor leading to the widespread use of Python in astronomy has been the development of  {\tt Astropy} \citep{astropy}, a project with the express goal of providing a single core package for astronomy in Python, as well as fostering interoperability between complementary Python astronomy packages.

Motivated by these trends, we present the first stable release of {\tt Halotools} (v0.2), an open-source,\footnote{{\tt Halotools} is licensed under a 3-clause BSD style license - see the licenses/LICENSE.rst file. Release {\tt v0.2} is archived with DOI 10.5281/zenodo.835898. While revising this paper for publication, code development has progressed such that the latest release is {\tt v0.5}, archived with DOI 10.5281/zenodo.835895.} object-oriented Python package for building and testing models of the galaxy--halo connection. {\tt Halotools} is community-driven, and already includes contributions from over a dozen scientists spread across numerous universities. Designed with high-speed performance in mind, the package generates mock observations of synthetic galaxy populations with sufficient speed to conduct expansive  Monte Carlo Markov Chain (MCMC) likelihood analyses over a diverse and highly customizable set of models.

{\tt Halotools} can be thought of in analogy to Boltzmann-solvers such as {\tt CMBFast} \citep{cmbfast}, {\tt CAMB} \citep{camb} and {\tt CLASS} \citep{class}. However, rather than generating $\Lambda{\rm CDM}$ predictions for Cosmic Microwave Background power spectra, {\tt Halotools} uses a halo catalog output of a cosmological simulation to predict large-scale structure observables such as galaxy clustering, galaxy--galaxy lensing, void abundance, redshift-space distortions, and related statistics. These predictions are forward modeled from a mock galaxy population: {\tt Halotools} makes no appeal whatsoever to simulation fitting functions \citep[e.g.,][]{sheth_tormen01,tinker05} that are commonly employed in conventional formulations of the galaxy--halo connection. As these and related fitting functions have well-established upper limits on their accuracy \citep[e.g.,][]{tinker08a}, our forward modeling approach makes {\tt Halotools} better suited to the needs of the precision cosmology program, limiting systematic uncertainty to the accuracy and cosmic variance of the underlying simulation.

{\tt Halotools} is an affiliated package\footnote{\url{http://www.astropy.org/affiliated}} of {\tt Astropy}. Affiliated packages are not part of the {\tt Astropy} core package, but extend the core with typically more domain-specific functionality while maintaining its engineering and interface standards. Since its inception, {\tt Halotools} has been written in fully public view on {\url{https://github.com/astropy/halotools}}, in the spirit of open science.

This paper is not intended to be viewed as code documentation, which is available online at {\url{http://halotools.readthedocs.io}}. Our aim here is to give a simple overview of the features of the package, and to serve as a standard reference for scientists using {\tt Halotools} to support their published work. We outline the basic structure of the code in \S\ref{section:overview}, describe the development workflow in \S\ref{section:development} and conclude in \S\ref{section:conclusion}.

\section{Package Overview}
\label{section:overview}

{\tt Halotools} is composed almost entirely in Python with syntax that is compatible with both 2.7 and 3.x versions of the language. Bounds-checking, exception-handling, and high-level control flow are always written in pure Python. Whenever possible, performance-critical functions are parallelized using Python's native {\tt multiprocessing} module. {\tt Halotools} relies heavily on vectorized functions in {\tt NumPy} as a core optimization strategy. However, in many cases there is simply no memory efficient way to vectorize a calculation, and it becomes necessary to write explicit loops over large numbers of points. In such situations, care is taken to pinpoint the specific part of the calculation that is the bottleneck; that section, and that section only, is written in {\tt Cython}.\footnote{Cython is a tool that compiles Python-like code into C code. See \url{http://cython.org} and \citet{cython} for further information.}

{\tt Halotools} is designed with a high degree of modularity, so that users can pick and choose the features that are suitable to their science applications. At the highest level, this modularity is reflected in the organization of the package into sub-packages. For {\tt Halotools} v0.2, there are three major sub-packages. The \sims sub-package described in \S\ref{subsection:sim_manager} is responsible for reducing ``raw" halo catalogs into efficiently organized fast-loading hdf5 files, and for creating and keeping track of a persistent memory of where the simulation data is stored on disk. The \emodels sub-package described in \S\ref{subsection:empirical_models} contains models of the galaxy--halo connection, as well as a flexible object-oriented platform for users to design their own models. The \mockobs sub-package described in \S\ref{subsection:mock_observables} contains a collection of functions that generate predictions for models in a manner that can be directly compared to astronomical observations. Many of the functions in \mockobs should also be of general use in the analysis of halo catalogs.

Although these sub-packages are designed to work together, each individual sub-package has entirely stand-alone functionality that is intended to be useful even in the absence of the others. For example, while {\tt Halotools} provides pre-processed halo catalogs that are science-ready as soon as they are downloaded, use of {\tt Halotools}-provided catalogs is entirely optional and the package works equally well with alternative halo catalogs provided and processed by the user. The \emodels sub-package can be used to populate mock galaxies into the halos of any cosmological simulation, where the populated halos could be identified by any algorithm. The functions in the \mockobs sub-package accept simple point-data as inputs, and so these functions could be used to generate observational predictions for semi-analytical models that otherwise have no connection to {\tt Halotools}. We outline each of these sub-packages in the sections below.

\subsection{Managing Simulation Data}
\label{subsection:sim_manager}

The end result of any algorithm for identifying dark matter halos in an N-body simulation is a catalog of tabular data storing the positions and properties of the halos (e.g., \citealt[][{\tt BDM}]{klypin_holtzman97}; \citealt[][{\tt SUBFIND}]{springel_etal01}; \citealt[][{\tt ROCKSTAR}]{behroozi_rockstar11}). One of the most tedious tasks in simulation analysis is the initial process of getting started with a halo catalog: reading large data files storing the halos (typically ASCII-formatted), making cuts on halos, adding additional columns, and storing the reduced and value-added catalog on disk for later use. In our experience with simulation analysis, one of the most common sources of errors comes from these initial bookkeeping exercises.

\begin{figure*}
\center
\caption{Loading a cached halo catalog into memory\label{code:cachedsim}}
\vspace{0.1in}
\fbox{\includegraphics[scale=0.82,clip]{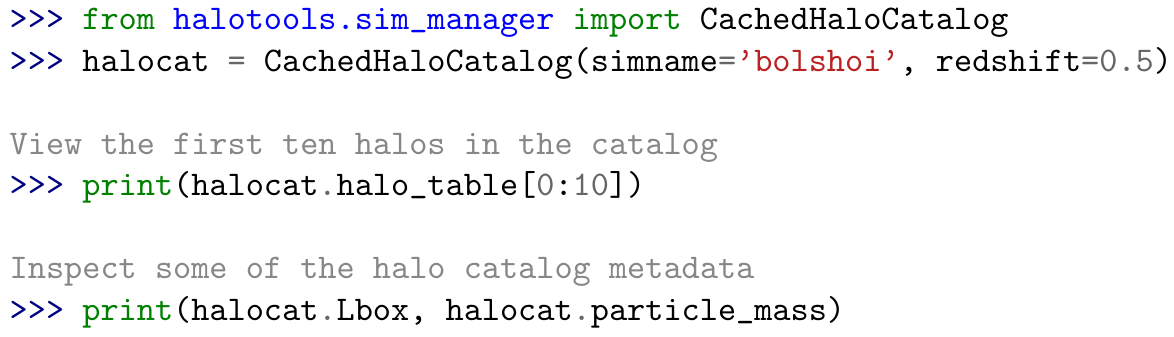}}
\end{figure*}

To help beginning users get started, the {\tt Halotools} developers manage a collection of pre-processed halo catalogs that are available for download either with the {\tt DownloadManager} class, or equivalently with the command-line script {\tt download\_additional\_halocat}. Through either download method, the catalogs are automatically cached and science-ready as soon as the download completes. {\tt Halotools} currently offers pre-processed halo catalogs for Rockstar-identified halos from four different simulations: {\tt bolshoi, bolshoi-planck, multidark} and {\tt consuelo}, for which snapshots at $z=0, 0.5, 1$ and $2$ are available \citep{gottloeber_klypin08, bolshoi_11, riebe_etal11, lasdamas}. A random downsampling of dark matter particles is also available to accompany each supported snapshot. The \sims sub-package structure also supports halo catalogs in light code format, although all pre-processed catalogs are single-snapshot data. 

In addition to these pre-processed catalogs, the \sims sub-package offers tools to process and format other halo catalogs so that users can get started with full-fledge halo catalog analysis from just a few lines of code.\footnote{All classes referred to in this section can be imported directly from the {\tt sim\_manager} sub-package.} The {\tt RockstarHlistReader} class allows users to quickly create a {\tt Halotools}-formatted catalog starting from the typical ASCII output of the {\tt Rockstar} halo-finder \citep{behroozi_rockstar11, rockstar_trees}. Users wishing to work with catalogs of halos identified by algorithms other than {\tt Rockstar} can use the {\tt TabularAsciiReader} class to initially process their ASCII data. Both readers are built around a convenient API that uses Python's native ``lazy evaluation" functionality to select on-the-fly only those columns and rows that are of interest, making these readers highly memory efficient.

All {\tt Halotools}-formatted catalogs are Python objects storing the halo catalog itself in the form of an {\tt Astropy} Table, and also storing some metadata about the simulated halos. In order to build an instance of a {\tt Halotools}-formatted catalog, a large collection of self-consistency checks about the halo data and metadata are performed,\footnote{For example, \sims verifies that all halo positions are confined to the simulation volume.} and an exception is raised if any inconsistency is detected. These checks are automatically carried out at the initial processing stage, and also every time the catalog is loaded into memory, to help ensure that the catalog is processed correctly and does not become corrupt over time.

The \sims sub-package allows users to cache their processed halo catalogs when they are saved to disk, creating the option to load their catalogs into memory with the simple and intuitive syntax shown in Figure \ref{code:cachedsim}. Cached simulations are stored in the form of an hdf5 file\footnote{\url{http://www.h5py.org}} \citep{hdf5}. This binary file format is fast-loading and permits metadata to be bound directly to the file in a transparent manner, so that the cached binary file is a self-expressive object. The {\tt HaloTableCache} class provides an object-oriented interface for managing the cache of simulations, but users are also free to work directly with the cache log, which is a simple, human-readable text file located in the {\tt Halotools} cache directory.

Using the {\tt Halotools} caching system is optional in every respect. Users who prefer their own system for managing simulated data are free to do so in whatever manner they wish; they need only pass the necessary halo data and metadata to the {\tt UserSuppliedHaloCatalog} class, and the full functionality of all sub-packages of {\tt Halotools} works with the resulting object instance.

\subsection{Empirical Models}
\label{subsection:empirical_models}

All {\tt Halotools} models of the galaxy--halo connection are contained in the \emodels sub-package. {\tt Halotools} models come in two categories: {\em composite models} and {\em component models}.
A {\em composite model} is a complete description of the mapping(s) between dark matter halos and all properties of their resident galaxy population. A composite model provides sufficient information to populate an ensemble of halos with a Monte Carlo realization of a galaxy population. All composite models are built from a collection of independently-defined {\em component models}. A component model provides a map between dark matter halos and a single property of the resident galaxy population. Example component models include the stellar-to-halo mass relation, a Navarro-Frenk-White (NFW) radial profile for the satellite distribution, or the halo mass-dependence of the quenched fraction. As of v0.2, the \emodels sub-package only contains models mapping snapshots of dark matter halos to galaxy populations at a fixed redshift. We elaborate upon the structure of {\tt Halotools} models below.

\subsubsection{Model styles}
\label{subsubsection:modelstyles}

{\tt Halotools} composite models come in two different types: HOD-style models and subhalo-based models. In HOD-style models, there is no relationship between the abundance of satellite galaxies in a host halo and the number of subhalos in that host halo. In these models, satellite abundance in each halo is determined by a Monte Carlo realization of some analytical model. Examples of this approach to the galaxy--halo connection include the HOD \citep{berlind02} and CLF \citep{yang03}, as well as extensions of these that include additional features such as color-dependence \citep{tinker_etal13}.

By contrast, in subhalo-based models there is a one-to-one correspondence between subhalos and satellite galaxies. In these models, each host halo in the simulation is connected to a single central galaxy, and each subhalo is connected to a single satellite. Examples include traditional abundance matching \citep{kravtsov04a,conroy06}, age matching \citep{hearin_etal13b}, and parameterized stellar-to-halo mass models  \citep{behroozi10, moster10}.

\subsubsection{Prebuilt models}
\label{subsubsection:prebuiltmodels}

\begin{figure*}
\center
\caption{Populating a mock galaxy catalog\label{code:mockpop}}
\vspace{0.1in}
\fbox{\includegraphics[scale=0.82,clip]{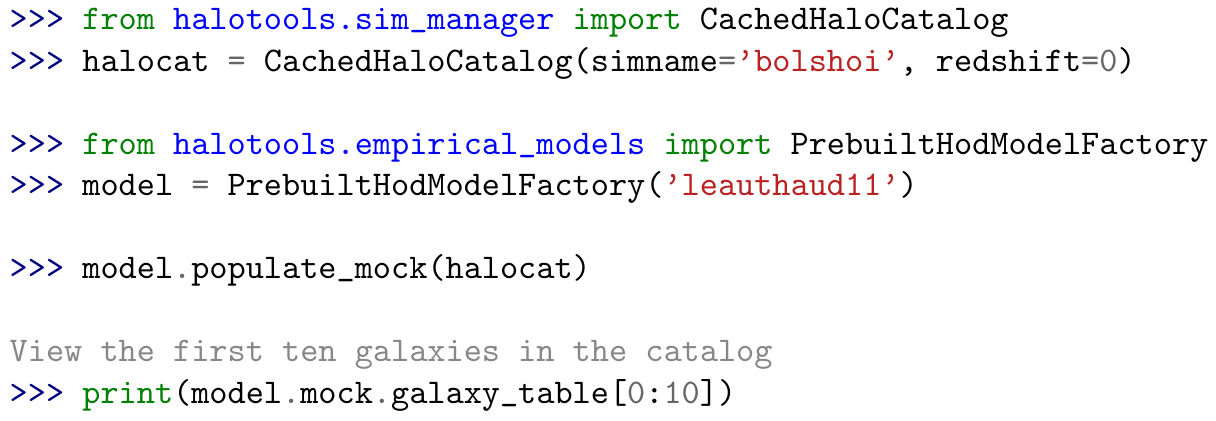}}
\end{figure*}

{\tt Halotools} ships with a handful of fully-formed prebuilt composite models, each of which has been designed around a model chosen from the literature. All pre-built models can directly populate a simulation with a mock catalog. Users need only choose the pre-built model and simulation snapshot that is appropriate for their science application, and can then immediately generate a Monte Carlo realization of the model. The syntax in Figure \ref{code:mockpop} shows how to populate the Bolshoi simulation at $z=0$ with an HOD-style model based on \citet{leauthaud11b}. All {\tt Halotools} models can populate halo catalogs with mock galaxies using this same syntax, regardless of the features of the model or the selected halo catalog.

Calling the {\tt populate\_mock} method the first time creates the {\tt mock} attribute of a model, and triggers a large amount of pre-processing that need only be done once. The algorithm used to repopulate mock catalogs takes advantage of this pre-processing, so that subsequent calls to {\tt model.mock.populate()} are far faster. For example, repopulation of the model shown in Figure \ref{code:mockpop} based on the \citet{leauthaud11b} model takes just a few hundred milliseconds on a modern laptop.

The behavior of all {\tt Halotools} models is controlled by the {\tt model.param\_dict} attribute, a Python dictionary storing the values of the model parameters. By changing the values of the parameters in the {\tt param\_dict}, users can generate alternate mock catalogs based on the updated parameter values. The process of varying {\tt param\_dict} values and repeatedly populating mock catalogs is the typical workflow in an MCMC-type analysis conducted with {\tt Halotools}.

\begin{figure}
\begin{center}
\includegraphics[width=8.3cm]{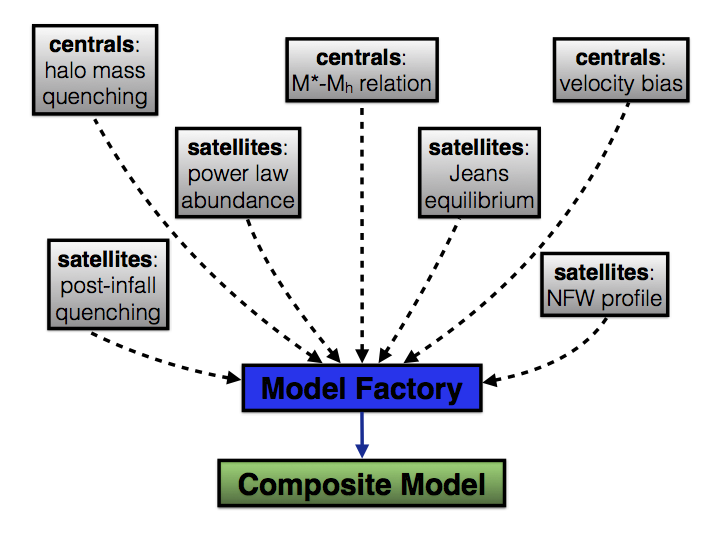}
\caption{
Cartoon example of how an HOD-style model can be built with the {\tt Halotools} factory design pattern; subhalo-based models can be built in a directly analogous fashion. Users select a set of {\em component model} features of their choosing, and compose them together into a {\em composite model} using the appropriate factory. Users wishing to quickly get up-and-running can instead select a prebuilt composite model that ships with the package.
}
\label{fig:modeling}
\end{center}
\end{figure}

\subsubsection{User-built models}
\label{subsubsection:userbuiltmodels}

The \emodels sub-package provides far more functionality than a simple set of prebuilt composite models that are ready to generate mock catalogs ``out of the box". {\tt Halotools} has special factory classes that allow users to build their own models connecting galaxies to the dark matter halos that host them. These factories are the foundation of the object-oriented platform that {\tt Halotools} users can exploit to design their own models of the galaxy--halo connection. This model-building platform is the centerpiece of the \emodels sub-package.

Users choose between a set of component models of the galaxy population and compose them together into a composite model using the appropriate {\tt Halotools} factory class; HOD-style models are built by the {\tt HodModelFactory} class, subhalo-based models are built by the {\tt SubhaloModelFactory}. Composing together different collections of components gives users a large amount of flexibility to construct highly complex models of galaxy evolution. There are no limits on the number of component models that can be chosen, nor on the number or kind of galaxy population(s) that make up the universe in the user-defined composite model. Figure \ref{fig:modeling} shows a cartoon example of the factory design of an HOD-style model.

In choosing component models, users are not restricted to the set of features that ship with the {\tt Halotools} package. Users are free to write their own component models and use the {\tt Halotools} factories to build the composite, to write just one new component model and include it in a collection of {\tt Halotools}-provided components, or anywhere in between. This way, users mostly interested in a specific feature of the galaxy population can focus exclusively on developing code for that one feature, and use existing {\tt Halotools} components to model the remaining features. The factory design pattern also makes it simple to swap out individual features while keeping all other model aspects identical, facilitating users to ask targeted science questions about galaxy evolution and answer these questions via direct computation.

\subsection{Mock Observations}
\label{subsection:mock_observables}

In the analysis of halo and (mock) galaxy catalogs, many of the same calculations are performed over and over again.
\bit
\item How many pairs of points are separated by some distance $r$?
\item What is the two-point correlation function of some sample of points?
\item What are the host halo masses of some sample of subhalos?
\item What is the local environmental density of some collection of galaxies?
\eit
It is common to calculate the answers to these and other similar questions in an MCMC-type analysis, when high-performance is paramount. Even outside of the context of likelihood analyses, the sheer size of present-day cosmological simulations presents a formidable computational challenge to evaluate such functions in a reasonable runtime. There is also the notorious complicating nuisance of properly accounting for the periodic boundary conditions of a simulation. Much research time has been spent by many different researchers writing their own private versions of these calculations, writing code that is not extensible as it was developed making hard assumptions that are only applicable to the immediate problem at hand.

\begin{figure*}
\begin{center}
\includegraphics[width=8.3cm]{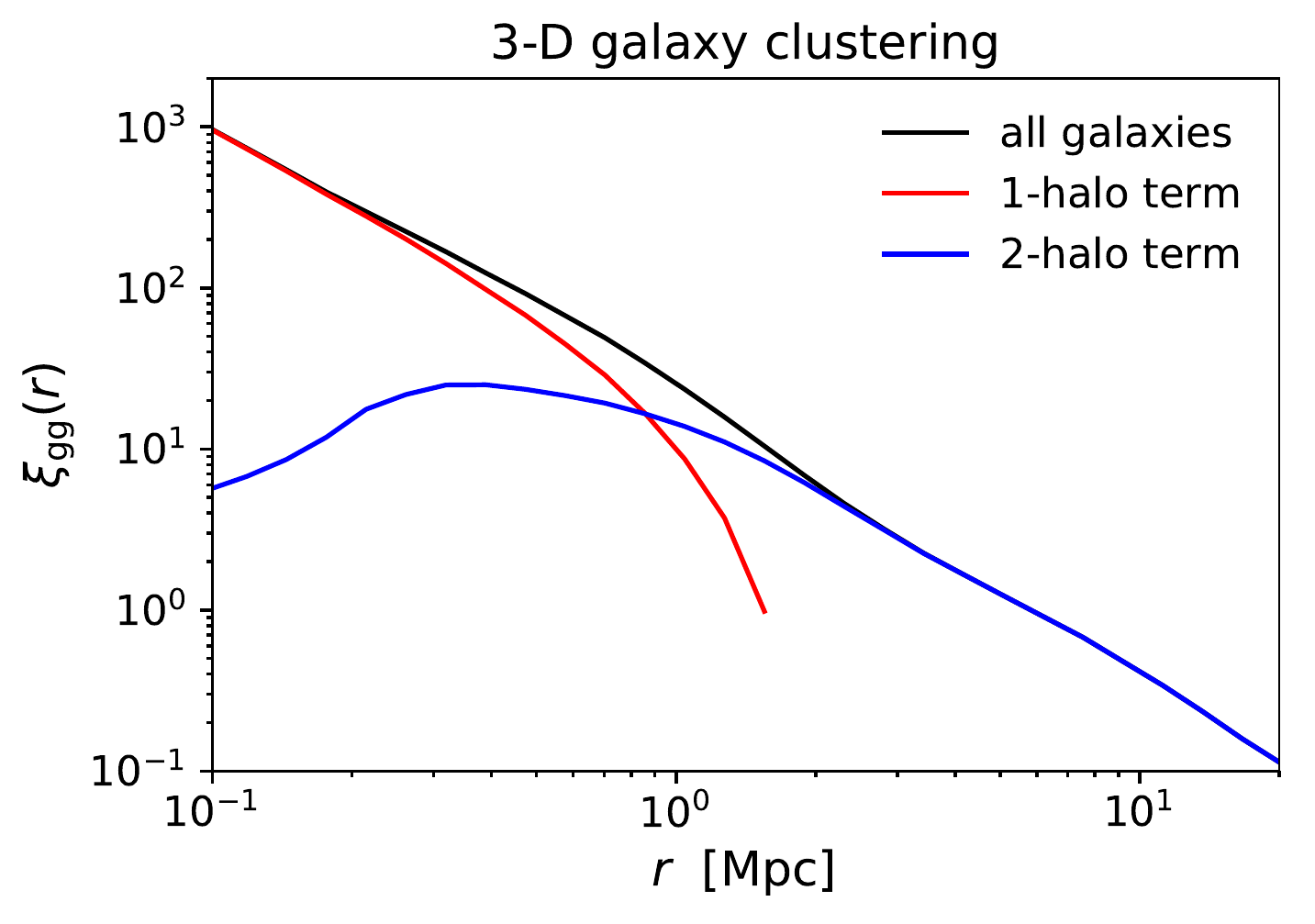}
\includegraphics[width=8.3cm]{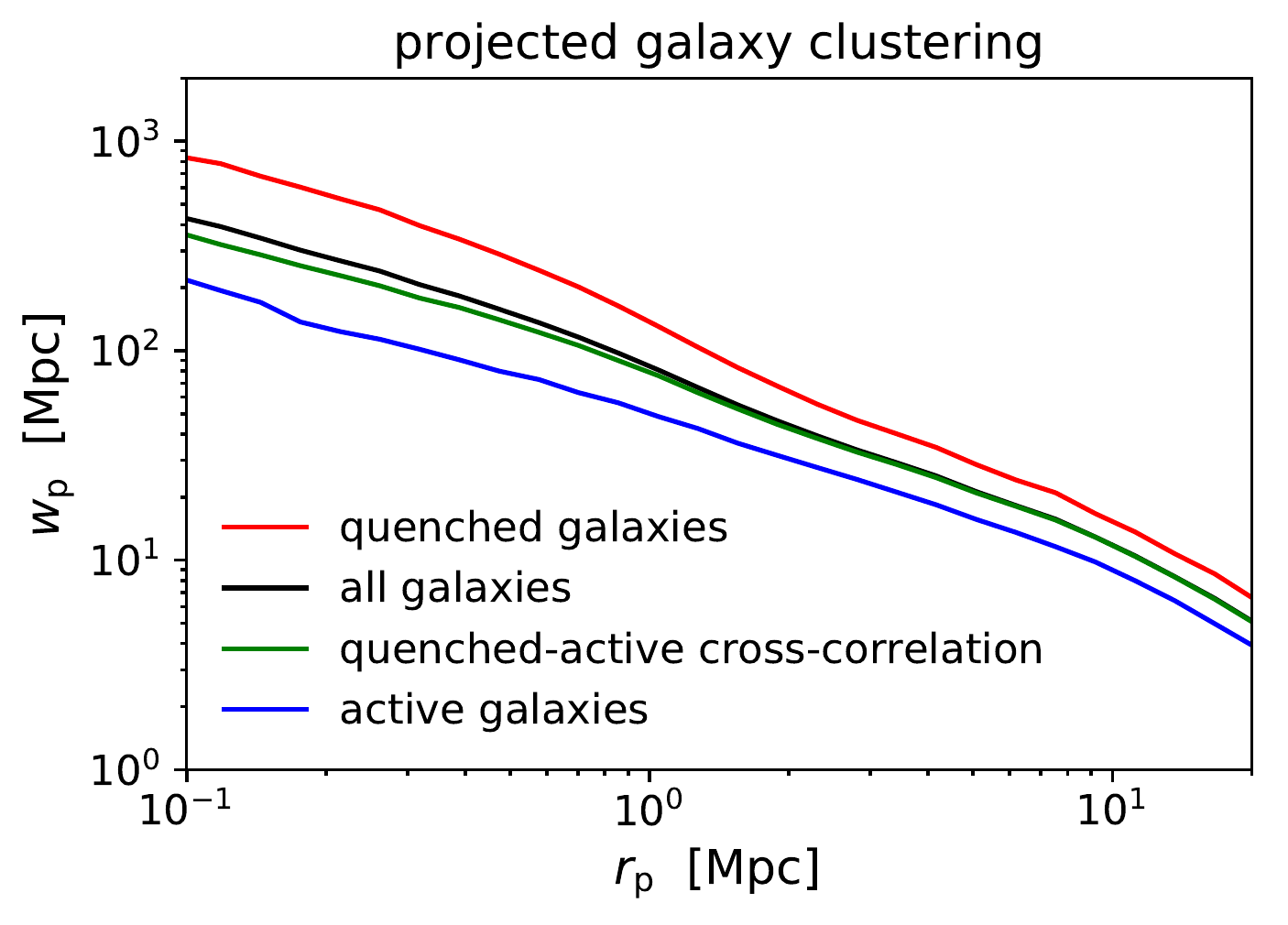}
\includegraphics[width=8.3cm]{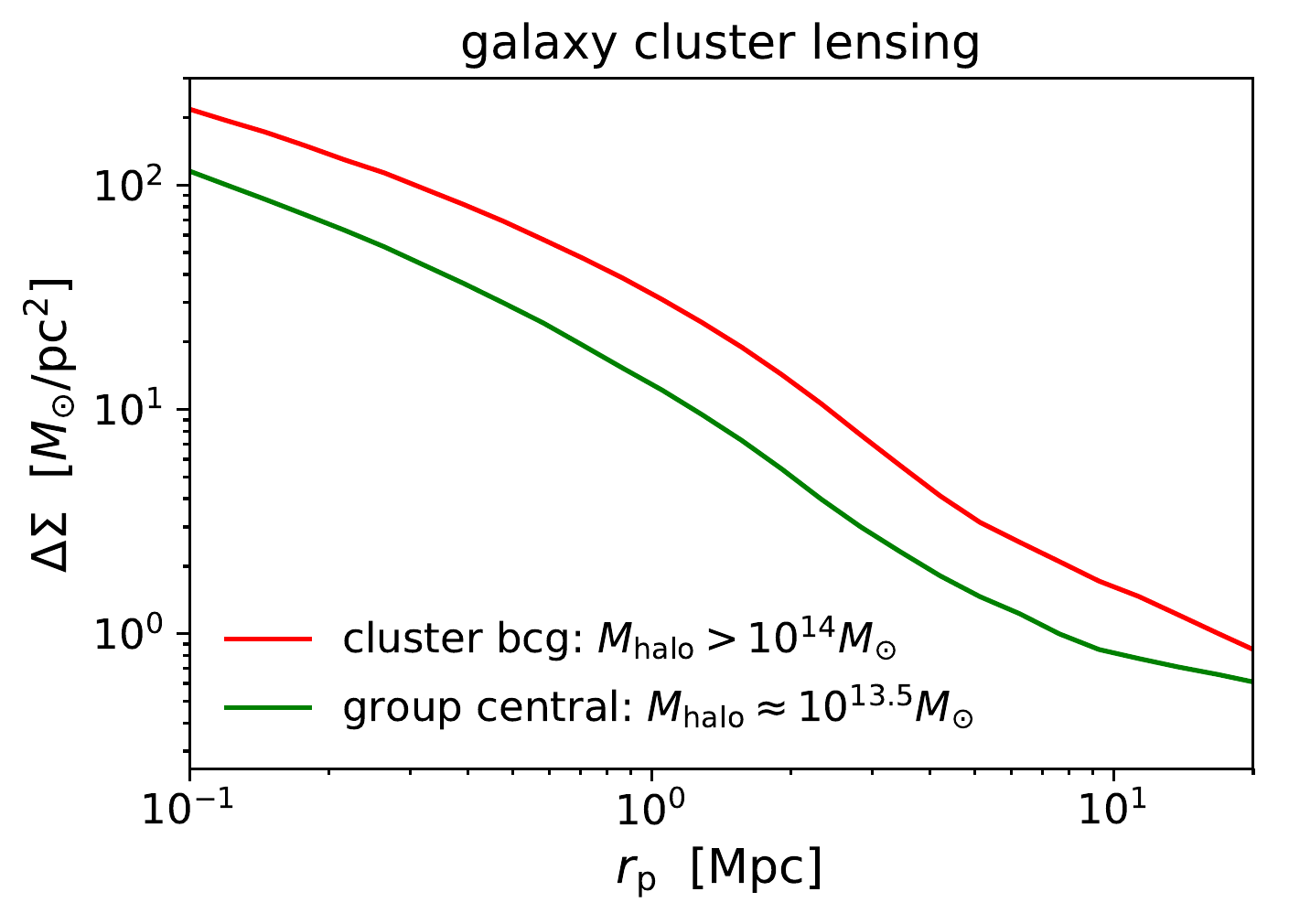}
\includegraphics[width=8.3cm]{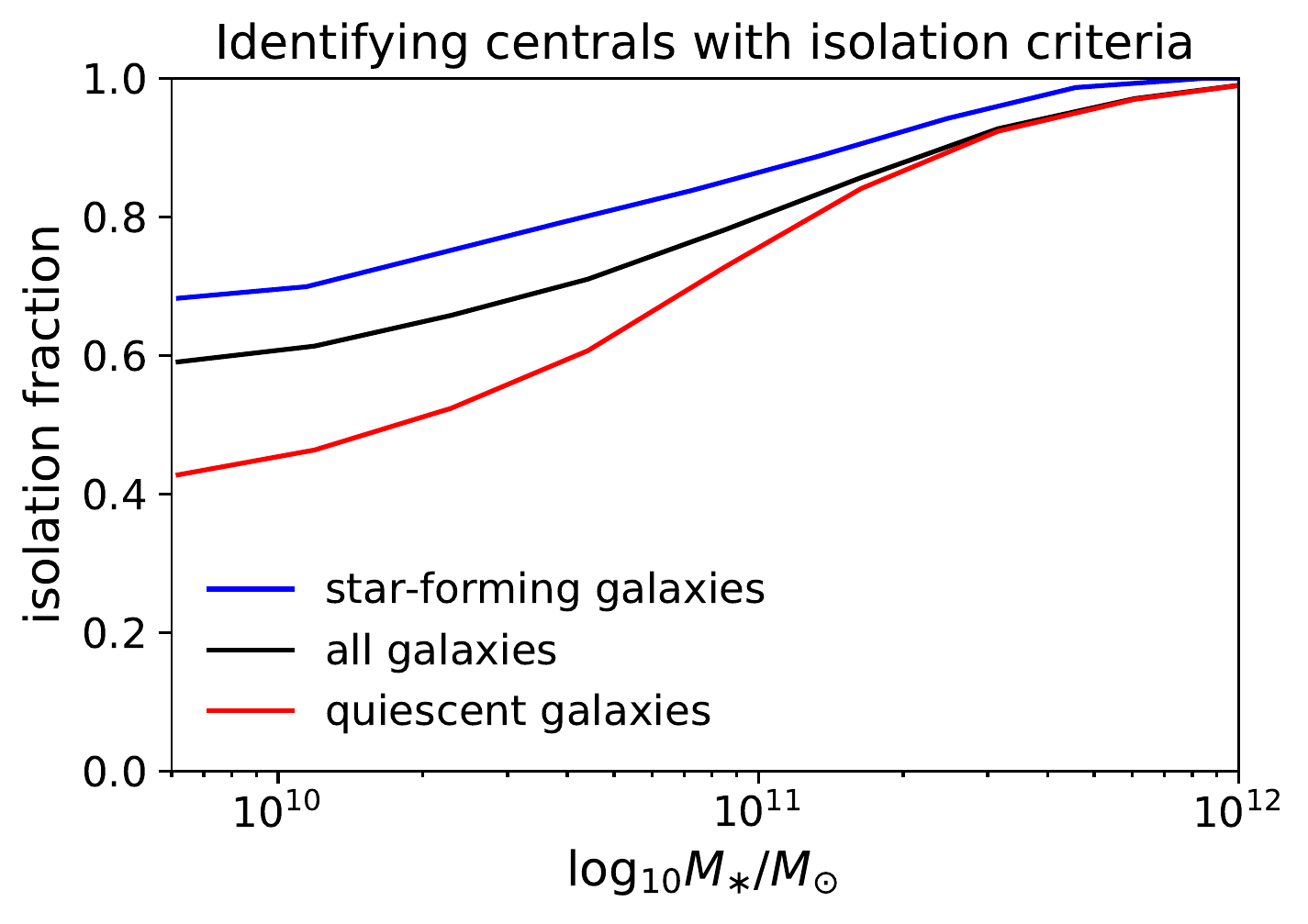}
\includegraphics[width=8.3cm]{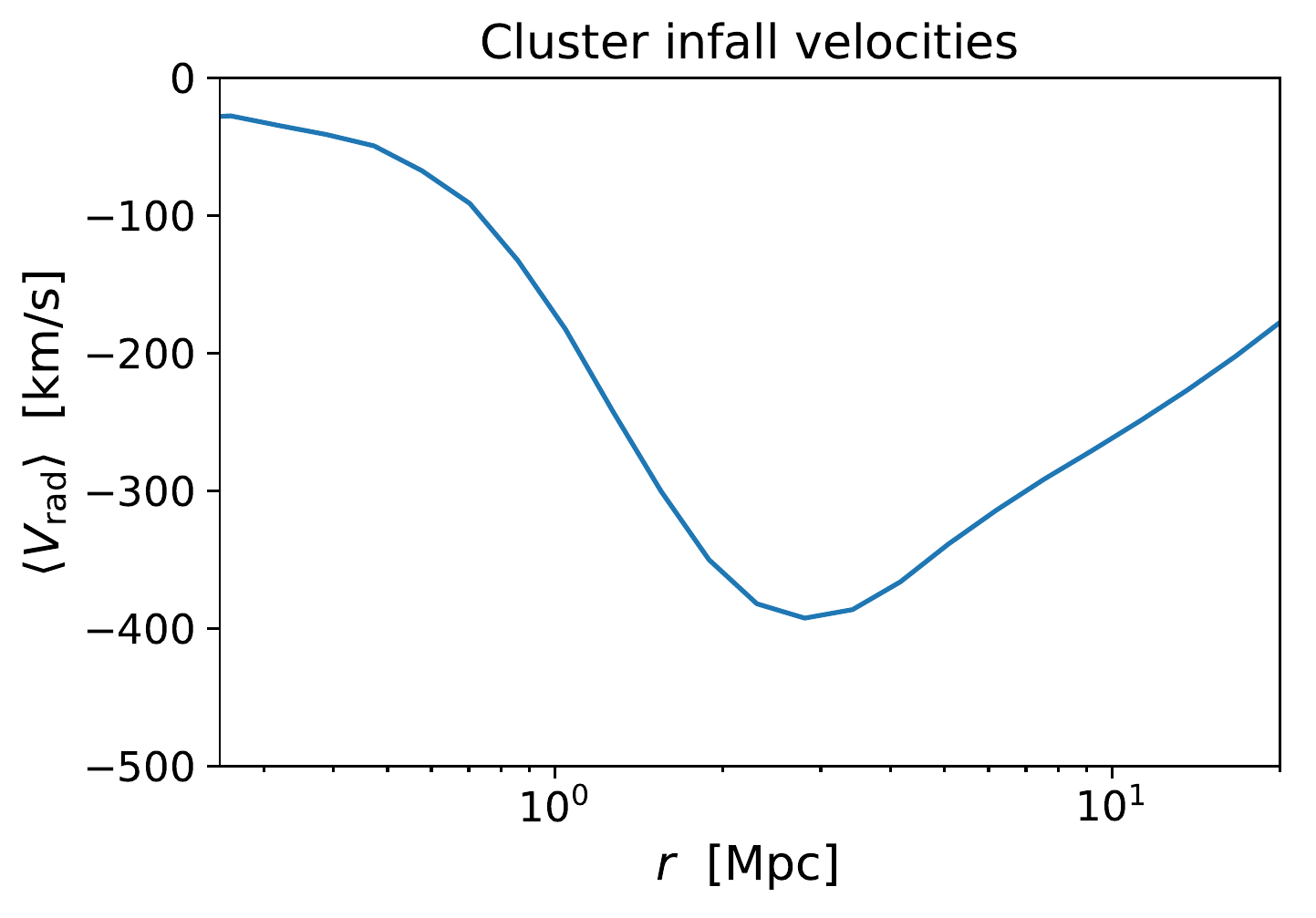}
\includegraphics[width=8.3cm]{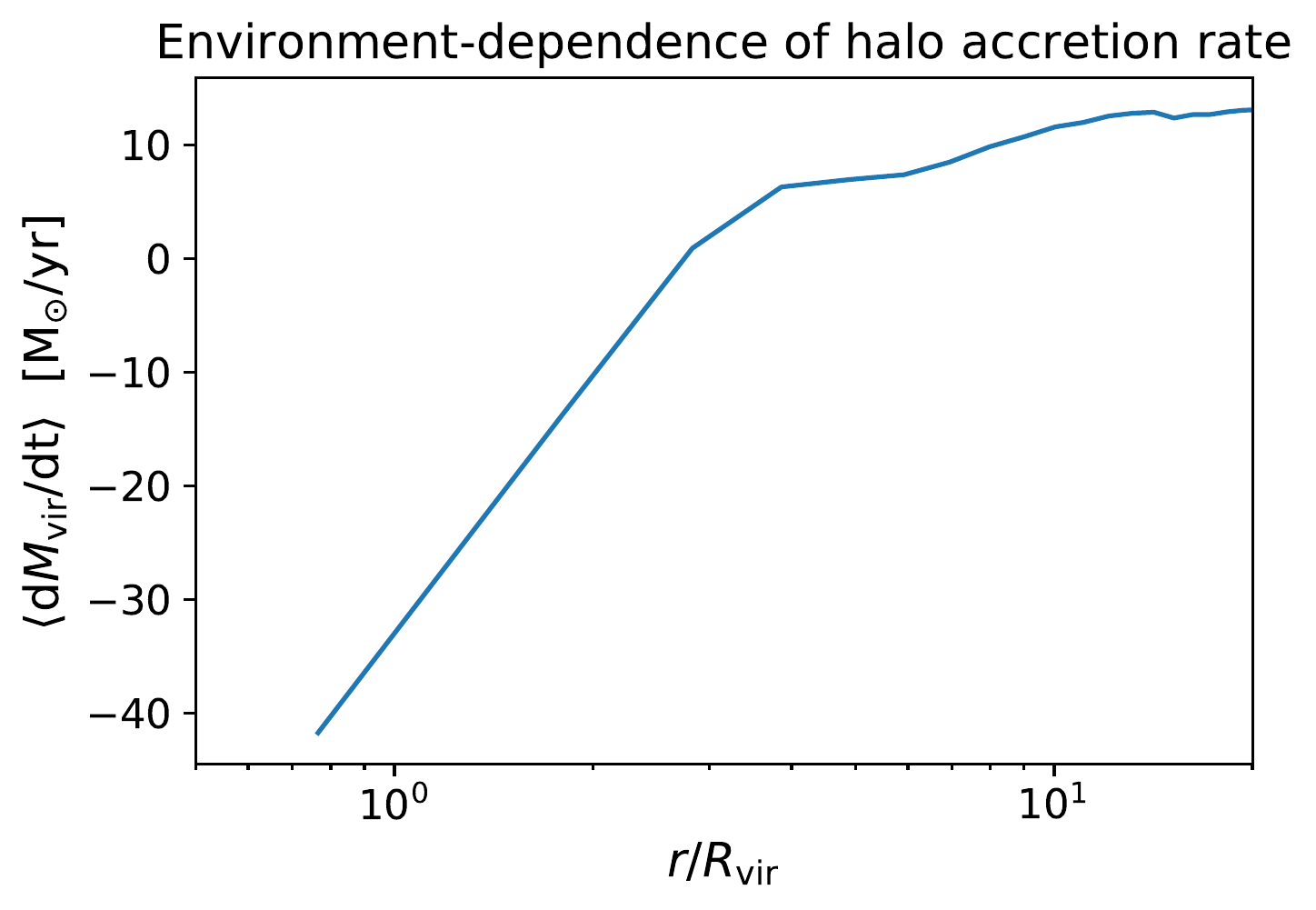}
\caption{
Six example calculations done with {\tt Halotools} demonstrating the diversity of the \mockobs sub-package. {\em Top left:} Three-dimensional correlation function of mock galaxies $\xi_{\rm gg}(r)$ split into contributions from pairs of galaxies occupying a common halo (1-halo term), and pairs in distinct halos (2-halo term). {\em Top right:} Projected correlation function $w_{\rm p}(\rproj)$ of star-forming and quiescent galaxies, as well as their cross-correlation. {\em Center left:} Galaxy lensing signal produced by galaxy clusters and galaxy groups. {\em Center right:} Fraction of galaxies that are considered ``isolated" according to whether there is a more massive companion within a cylinder of projected radius $500$kpc and a velocity difference of $500$km/s. {\em Bottom left:} Mean pairwise radial velocity of galaxies in the neighborhood of a cluster BCG. {\em Bottom right:} Mean mass accretion rate of lower-mass subhalos as a function of cluster-centric distance $r/R_{\rm vir}.$
}
\label{fig:mockobs}
\end{center}
\end{figure*}

The {\tt mock\_observables} sub-package is designed to remedy this situation. This sub-package contains a large collection of functions that are commonly encountered when analyzing halo and galaxy catalogs, including:

\bit
\item The many variations of two-point correlation functions,
\bit
\item three-dimensional correlation function $\xi(r),$
\item redshift-space correlation function $\xi(\rproj, \pi),$
\item projected correlation function $w_{\rm p}(\rproj),$
\item projected surface density $\Delta\Sigma(r_{\rm p})$ (aka galaxy--galaxy lensing),
\item RSD multipoles $\xi_{\ell}(s).$
\eit
\item marked correlation functions $\mathcal{M}(r),$
\item friends-of-friends group identification,
\item {\em group aggregation} calculations, e.g., calculating the total stellar mass of galaxies of a common group $M_{\ast}^{\rm tot},$
\item {\em isolation criteria}, e.g., identifying those galaxies without a more massive companion inside some search radius,
\item pairwise velocity statistics, e.g, the line-of-sight velocity dispersion as a function of projected distance $\sigma_{\rm los}(\rproj),$
\item void probability function $P_{\rm void}(r),$
\item etc.
\eit
The \mockobs sub-package contains heavily optimized implementations of all the above functions, as well as a variety of others. Figure \ref{fig:mockobs} provides a visual demonstration of the diversity of the available options.

Every function in \mockobs has a stable, user-friendly API that is consistently applied across the package. The docstring of all functions contains an explicit example of how to call the function, and in many cases there is a step-by-step tutorial in the documentation showing how the function might be used in a typical analysis. Considerable effort has been taken to write \mockobs to be modular, so that users can easily borrow the algorithm patterns to write their own variation on the provided calculations.

\section{Package Development}
\label{section:development}

\subsection{GitHub workflow}
\label{subsection:githubworkflow}

{\tt Halotools} has been developed fully in the open since the inception of the project. Version control for the code base is managed using {\tt git}\footnote{\url{http://git-scm.com}}, and the public version of the code is hosted on GitHub\footnote{\url{http://www.github.com}}. The latest stable version of the code can be installed via {\tt pip install halotools} or via {\tt conda install -c astropy halotools}, but at any given time the {\tt master} branch of the code on \url{https://github.com/astropy/halotools} may have features and performance enhancements that are being prepared for the next release. A concerted effort is made to ensure that only thoroughly tested and documented code appears in the public {\tt master} branch, though {\tt Halotools} users should be aware of the distinction between the bleeding edge version in {\tt master} and the official release version available through {\tt pip} or {\tt conda}.

Development of the code is managed with a {\em Fork and Pull} workflow. Briefly, code development begins by creating a personal {\em fork} of the main repository on GitHub. Developers then work only on the code in their fork. In order to incorporate a change to the main repository, it is necessary to issue a {\em Pull Request} to the {\tt master} branch. The version of the code in the Pull Request is then tested and reviewed by the {\tt Halotools} developers before it is eligible to be merged into {\tt master}.

\subsection{Automated testing}
\label{subsection:testing}

{\tt Halotools} includes hundreds of unit tests that are incorporated into the package via the {\tt py.test} framework.\footnote{\url{http://pytest.org}} These tests are typically small blocks of code that test a specific feature of a specific function. The purpose of the testing framework is both to verify scientific correctness and also to enforce that the API of the package remains stable. We also use {\em continuous integration}, a term referring to the automated process of running the entire test suite in a variety of different system configurations (e.g., with different releases of {\tt NumPy} and {\tt Astropy} installed, or different versions of the Python language). Each time any Pull Request is submitted to the {\tt master} branch of the code on GitHub, the proposed new version of the code is automatically copied to a variety of virtual environments and the entire test suite is run repeatedly in each environment configuration. The Pull Request will not be merged into {\tt master} unless the entire test suite passes in all environment configurations. We use {\tt Travis}\footnote{\url{https://travis-ci.org}} for continuous integration in Unix environments such as Linux and Mac OS X, and {\tt AppVeyor}\footnote{\url{https://www.appveyor.com}} for Windows environments.

Pull Requests to the {\tt master} branch are additionally subject to a requirement enforced by {\tt Coveralls}.\footnote{\url{https://coveralls.io}} This service performs a static analysis on the {\tt Halotools} code base and determines the portions of the code that are covered by the test suite, making it straightforward to identify logical branches whose behavior remains to be tested. {\tt Coveralls} issues a report for the fraction of the code base that is covered by the test suite; if the returned value of this fraction is smaller than the coverage fraction of the current version of {\tt master}, the Pull Request is not accepted. This ensures that test coverage can only improve as the code evolves and new features are added.

Any time a bug is found in the code, either by {\tt Halotools} developers or users, a GitHub Issue is opened calling public attention to the problem. When the {\tt Halotools} developers have resolved the problem, a corresponding {\em regression test} becomes part of the code base. The regression test explicitly demonstrates the specific source of the problem, and contains a hyperlink to the corresponding GitHub issue. The test will fail when executed from the version of the code that had the problem, and will pass in the version with the fix. Regression testing helps to make it transparent how the bug was resolved and protects against the same bug from creeping back into the repository as the code evolves.

\subsection{Documentation}
\label{subsection:documentation}

Documentation of the code base is generated with {\tt Sphinx}\footnote{\url{http://www.sphinx-doc.org}} and is hosted on {\tt Read the Docs}\footnote{\url{ https://readthedocs.org}} (RTD) at \url{http://halotools.readthedocs.io}. The public repository \url{https://github.com/astropy/halotools} has a webhook set up so that whenever there is a change to the {\tt master} branch, the documentation is automatically rebuilt to reflect the most up-to-date version of {\tt master}.\footnote{In fact, there are two versions of documentation available, {\em latest} and {\em stable}, and one can choose on RTD which one to read.}

Every user-facing class, method and function in {\tt Halotools} has a docstring describing its general purpose, its inputs and outputs, and also providing an explicit example usage. Each such example usage appearing in a docstring also serves as a {\em doctest}. A doctest is a code fragment appearing in the documentation that 1.~demonstrates an example call to a function, and 2.~executes as part of the test suite. Thus if the API of a function changes but the documentation is not updated reflect this change, this triggers a test failure that must be resolved prior to merging the modified code into {\tt master}. Doctests help ensure that as {\tt Halotools} evolves, the code still behaves as the documentation says it does.

Docstrings for many functions with complex behavior come with a hyperlink to a separate section of the documentation in which mathematical derivations and algorithm notes are provided. The documentation also includes a large number of step-by-step tutorials and example analyses. The goal of these tutorials is more than simple code demonstration: the tutorials are intended to be a pedagogical tool illustrating how to analyze simulations and study models of the galaxy--halo connection in an efficient and reproducible manner.

\section{Conclusion}
\label{section:conclusion}

We have presented the first stable release of the {\tt Halotools} package (v0.2), a specialized Python package for building and testing models of the galaxy--halo connection, and analyzing catalogs of dark matter halos. The core functionality of the package includes:

\bit
\item Fast generation of synthetic galaxy populations using HODs, abundance matching, and related methods.
\item Efficient algorithms for calculating galaxy clustering, lensing, z-space distortions, and other astronomical statistics.
\item A modular, object-oriented framework for designing galaxy evolution models.
\item End-to-end support for reducing halo catalogs and caching them as fast-loading hdf5 files.
\eit

We offer several examples below of typical use-cases for which {\tt Halotools} was designed. This list is not intended to be complete, but simply to illustrate the full-featured nature of the package, which includes functionality supporting both galaxy evolution science as well as cosmological simulation analysis. The documentation hosted at \url{https://halotools.readthedocs.io} contains extensive tutorials and step-by-step guides that can be used as examples for how to carry out each of the example studies below and more.

\ben
\item Constrain traditional HOD model parameters with an MCMC-type analysis of observational measurements of projected galaxy clustering $w_{\rm p}(\rproj).$
\item Build a novel empirical model of galaxy morphology and derive MCMC-type constraints on its parameters using non-traditional large-scale structure measurements such as the morphology-marked correlation function $\mathcal{M}(\rproj).$
\item Conduct a study of the connection between the radial profiles of dark matter subhalos and the mass accretion history of their parent halo.
\item Build mock catalogs of galaxies that include forward-models of observational systematics such as intrinsic alignments and/or deblending errors in galaxy shapes.
\een

We would like to conclude this paper with an invitation. As can be seen from the public record of the package development on GitHub, contributions of all kinds are warmly welcomed. This could include submitting a Pull Request of a novel feature that has been developed for a scientific publication, or simply submitting a bug report or requesting documentation clarification. It is the hope of the development team that scientists who find the package useful in their own work will also use {\tt Halotools} as an outlet to share their expertise in large-scale structure in a way that can benefit the wider astronomical community.


\section{acknowledgments}

APH would like to express profound gratitude to the Yale Center for Astronomy and Astrophysics, and Meg Urry in particular, for their carte blanche support for this long-term and extremely labor-intensive project. Without the creative freedom offered by the YCAA Prize fellowship, there would be no {\tt Halotools}.

Work at Argonne National Laboratory was supported under U.S. Department of Energy contract DE-AC02-06CH11357.

APH gives special thanks to Jim Ford for every last groove of {\em Harlan County}. APH is grateful to Nikhil Padmanabhan and Frank van den Bosch for invaluable and continuing guidance. Thanks to Doug Watson and Matt Becker for productive discussions at a formative stage of {\tt Halotools} development, and Yu Feng for help in optimizing the \mockobs sub-package and comments on an early draft of this paper. Thanks to Andrey Kravtsov for help with establishing compatibility in Windows environments.

DC would like to thank Nikhil Padmanabhan and Frank van den Bosch for their patience and support while developing this package.

We thank the {\tt Astropy} developers for the package-template, which has been critical to the entire process of {\tt Halotools} development.
We would like to thank the NumPy, SciPy, IPython, Matplotlib, GitHub, Read the Docs, Travis, AppVeyor and Coveralls communities for their extremely useful free software products. We thank all {\tt Halotools} users who have patiently helped improve the package by providing feedback and critiques from the project's alpha-stages to the present.

A portion of this work was also supported by the National Science Foundation under grant PHYS-1066293 and the hospitality of the Aspen Center for Physics. This work was also partially funded by the U.S. National Science Foundation under grant AST 1517563. Support for EJT was provided by NASA through Hubble Fellowship grants \#51316.01 awarded by the Space Telescope Science Institute, which is operated by the Association of Universities for Research in Astronomy, Inc., for NASA, under contract NAS 5-26555.  PB was supported through program number HST-HF2-51353.001-A, provided by NASA through a Hubble Fellowship grant from STScI, which is operated by the Association of Universities for Research in Astronomy, Incorporated, under NASA contract NAS5-26555. NJG is funded by NSF grant ACI-1535651 as well as by the Gordon and Betty Moore Foundation's Data-Driven Discovery Initiative through Grant GBMF4651. EJ is supported by Fermi Research Alliance, LLC under the U.S. Department of Energy under contract No. DEAC02-07CH11359. SM is supported by Grant-in-Aid for Scientific Research from the JSPS Promotion of Science (15K17600, 16H01089).

\bibliographystyle{yahapj}
\bibliography{./halotools}

\end{document}